\documentclass{article}
\newcommand{\ft}[2]{{\textstyle\frac{#1}{#2}}}
\def\der{\partial}
\newcommand{\mscr}[1]{\mbox{\scriptsize #1}}
\usepackage{fortschritte}
\def\beq{\begin{equation}}                     %
\def\eeq{\end{equation}}                       %
\def\bea{\begin{eqnarray}}                     
\def\eea{\end{eqnarray}}                       
\begin {document}                 
\def\titleline{
%
%
%
Topological Transitions and 
Enhan\c{c}on-like Geometries in Calabi-Yau Compactifications
of M-Theory
}
\def\email_speaker{
{\tt 
%
%
mohaupt@tpi.uni-jena.de
}}
\def\authors{
%
%
%
%
%
T.~Mohaupt\1ad\sp                           
}
\def\addresses{
%
%
%
%
\1ad                                        %
Theoretisch-Physikalisches Institut,
Friedrich-Schiller-Universit\"at Jena, \\
Max-Wien Platz 1, D-07743 Jena, Germany \\
}
\def\abstracttext{
%
%
We study the impact of topological phase transitions
of the internal Calabi-Yau threefold on the space-time
geometry of five-dimensional extremal black holes and black strings. For
flop transitions and $SU(2)$ gauge symmetry enhancement we
show that solutions can always be continued and that the
behaviour of metric, gauge fields and scalars can be
characterized in a model independent way. Then we look
at supersymmetric solutions which describe naked
singularities rather than geometries with a horizon.
For black strings we show that the solution cannot become
singular as long as the scalar fields take values inside
the K\"ahler cone.
For black holes we establish the same result for 
the
elliptic fibrations over the Hirzebruch surfaces
${\cal F}_0, {\cal F}_1, {\cal F}_2$.
These three models exhibit a behaviour similar to the enhan\c{c}on,
since one runs into $SU(2)$ enhancement before reaching
the apparent singularity. Using the proper continuation 
inside the enhan\c{c}on radius one finds that the solution
is regular.
}
\large
\makefront

M-theory generalizes quantum field theory by the introduction of 
extended objects: strings and branes. This leads to surprising
new phenomena, which include topological phase transitions and
the smoothing or excision of space-time singularities. The 
enhan\c{c}on \cite{ClifEtAl} and similar phenomena in Horava-Witten 
theory \cite{KalMohShm} 
show that degenerations in the internal space and the 
absence of space-time singularities are closely related.
In this paper we briefly summarize new results along these
lines.\footnote{A more datailed account will be given in 
a separate publication.}


The compactification of eleven-dimensional supergravity
on a Calabi-Yau threefold yields minimal five-dimensional
supergravity coupled to $n_V = h_{1,1}-1$ abelian vector
multiplets and $n_H = h_{2,1} +1$ hypermultiplets. The 
hypermultiplets are trivial in the space-time geometries
considered here. The couplings of vector multiplets
are encoded in the so-called very special geometry of the
scalar manifold, which is a cubic hypersurface  \cite{GST,dWvP}
\beq
{\cal V}(X) = \ft16 C_{IJK} X^I X^J X^K = 1 \;,\;\;\;
I,J,K = 0, \ldots n_V \;.
\label{Prepot}
\eeq
The function ${\cal V}(X)$ is called the prepotential, and
all couplings of the vector multiplets can be computed from it.
Thus the Lagrangian is fully determined by the (locally)
constant, real symmetric tensor $C_{IJK}$.

The electric BPS solutions of this theory take the following
form \cite{ChaSab}:
\bea
ds^2 &=& - {\cal V}(Y)^{-4/3} dt^2 + {\cal V}(Y)^{2/3} dx_{(4)}^2 
\nonumber \\
F^I_{tm} &=& - \der_{m} \left(  {\cal V}(Y)^{-1} Y^I \right) \;,
\;\;\; \phi^x = {Y^x}/{Y^0} \;. 
\label{BlackHole}
\eea
The solution is parametrized in terms of rescaled scalar 
fields $Y^I = a X^I$, where $a^3 = \ft16 C_{IJK} Y^I Y^J Y^K=
{\cal V}(Y)$. This is convenient because all space-time
quantities, i.e., the metric, the gauge fields 
$F^I_{\mu \nu}$ fields and unconstrained scalars $\phi^x$ are 
functions of the $Y^I$. Note that the $Y^I$ (or,
equivalently, $X^I$) are subject to one constraint. The
unconstrained scalars $\phi^x$, $x =1, \ldots, n_V$ can
be obtained by solving (\ref{Prepot}). Above we used
'special coordinates' $\phi^x = X^x/X^0 = Y^x/Y^0$.

The scalars $Y^I$ have to satisfy the equation
$\Delta \left(  C_{IJK} Y^J Y^K \right) = 0$, where
$\Delta$ is the (flat) Laplacian with respect to the four spatial
coordinates. The solution can be expressed
in terms of $n_V+1$ harmonic functions $H_I$ by solving the  
so-called generalized stabilization equations \cite{ChaSab}:
\beq
C_{IJK} Y^J Y^K = 2 H_I \;.
\label{FlowEq}
\eeq
This is a coupled system of quadratic equations for which
no closed solution exists, though one knows explicit solutions
for specific choices of $C_{IJK}$. We will only consider
single-centered harmonic functions $H_I = c_I + \ft{q_I}{r^2}$,
where $r$ is the transverse radial coordinate. Thus the solution
depends on $2(n_V + 1)$ parameters: whereas the $c_I$
determine the values of the scalars at infinity (one parameter
is fixed by normalizing the metric), the $q_I$ are the electric
charges carried by the solution. 

Not all such solutions describe black holes. 
To exhibit potential singularities we look
at the Ricci scalar:
\beq
R = - \frac{6 a a' + 4 r (a')^2 + 2 r a a'' }{r a^4} \;,
\label{Rscalar}
\eeq
where $a$ was defined above and $a'= \ft{da}{dr}$, etc. 
As discussed in \cite{ChaSab} the solution has a regular 
horizon isometric to (a slice of) $AdS^2 \times S^3$ if
the number of Killing spinors doubles at $r=0$. In this case
the values of the scalars at $r=0$ are fixed in terms of the
charges. This is the celebrated supersymmetric attractor 
mechanism \cite{FKS}.

We will focus here on the much less studied question whether
naked singularities can occure at $r=r_{\star} >0$, i.e.,
before reaching a horizon. By inspection of (\ref{Rscalar})
one finds that this happens if either $a(r)$ vanishes at 
$r_{\star}$, or if $a'(r)$ or $a''(r)$ diverge.\footnote{The analysis of 
other curvature invariants does not change the discussion.}
The presence or absence of such a behaviour imposes 
inequalities (and not equalities) on the parameters $c_I,q_I$: 
singular and regular solutions are equally generic.

We will argue later that the solutions with naked singularities
are unphysical in M-theory. But first we have to consider the
global structure of the scalar manifold. If our five-dimensional
supergravity theory has been obtained by compactification
of eleven-dimensional supergravity on a Calabi-Yau 
threefold \cite{Reduction11/5},
then all its data are determined in terms of topological 
properties of the internal space. In particular, the 
constants $C_{IJK}$ are the
triple-intersection numbers of the Calabi-Yau space, the scalars
parametrize deformations of its K\"ahler structure, and the
scalar manifold is a hypersurface in the K\"ahler cone.\footnote{The
scalar corresponding to the total volume of the internal space
sits in a hypermultiplet and is constant for our solutions.}

The K\"ahler cone has boundaries, which correspond to singular
Calabi-Yau spaces. 
The generic singularities and their physical interpretation were
explained in \cite{Wit}: 
\begin{enumerate}
\item
the collapse of a two-cycle to a 
zero-cycle is related to a flop transition. In physical terms
one gets massless charged hypermultiplets at the boundary. One
can add a new K\"ahler cone to obtain the extended K\"ahler cone.
\item
the collapse of a four-cycle to a two-cycle yields two charged
vector multiplets which enhance a $U(1)$ factor of the gauge group
to $SU(2)$. 
One can continue the moduli beyond this boundary, but
the new K\"ahler cone is a gauge equivalent copy (obtained by
an elementary transformation) of the original cone \cite{KatMorPle,MohZag}. 
One can
either work with the extended parameter space (the extended
movable cone) or restrict oneself to one cone, as we will see
below.
\item
the collapse of a four-cycle to a zero-cycle results in 
tensionless strings. In this case no continuation is possible. 
\end{enumerate}

In the following we will study the cases (1) and (2), where a
continuation is possible.
We assume that the K\"ahler cone can
be parametrized by $Y^I > 0$, where $Y^I$ is proportional
to the volume of a holomorphic curve which collapses on the
boundary $Y^I=0$.\footnote{This is always possible
for toric Calabi-Yau threefolds, and, more generally, if there
is a basis for the homological two-cycles consisting of cycles which are
'nef' (numerically effective). In this basis the $C_{IJK}$ are non-negative.
We thank Albrecht Klemm for explaining this to us.} 
Then the continuation of the prepotential beyond the boundary
$Y^b=0$, $b \in \{ I \}$ is \cite{Wit,KatMorPle,MohZag}:
\beq
\tilde{\cal V}(Y) = \ft16 \tilde{C}_{IJK} Y^I Y^J Y^K \;,
\eeq
where
\beq
\tilde{C}_{IJK} = C_{IJK} + ( \delta  n_V - \delta n_H )
\delta_{Ib} \delta_{Jb} \delta_{Kb} \;.
\eeq
Here $\delta n_V$ and $\delta n_H$ are the numbers of 
charged vector and hypermultiplets which become massless
at $Y^b = 0$.\footnote{They become massive again
for negative $Y^b$.}

Now suppose that we have found a solution of (\ref{FlowEq})
such that $Y^I(r) > 0$ for $r > r_{\star}$ and 
$Y^b(r_{\star})=0$. In order to continue the solution to
$r<r_{\star}$ we must take into account that the coefficients
$C_{IJK}$ change when $Y^b$ changes sign. Therefore the
interior solution, where $r<r_{\star}$, is obtained by solving
\bea
\tilde{C}_{IJK} Y^J Y^K = 2 \tilde{H}_I \;.
\label{FlowEq2}
\eea
Then one has to glue the two branches of the 
solution at $r=r_{\star}$.
Note that it is not clear a priori whether the interior solution
depends on the same harmonic functions as the exterior solution.
Since the general solution depends on $2(n_V +1)$ parameters,
we need to impose $2(n_V +1)$ conditions to obtain a unique
continuation. The natural matching conditions are:
\bea
Y^I_{\mscr{(ext)}}(r_{\star}+) = 
Y^I_{\mscr{(int)}}(r_{\star}-) \;,\;\;\;
F^I_{tr \;\mscr{(ext)} }(r_{\star}+) =
F^I_{tr \;\mscr{(int)} }(r_{\star}-) \;,
\eea
which require that scalars and gauge fields are continuous at $r_{\star}$.
This implies that the metric is continuous as well, since it is
an algebraic
function of ${\cal V}(Y)$. Note that although
the coefficients $C_{IJK}$ jump at $Y^b=0$, the function
${\cal V}(Y)$ is continuous: the only
coefficient which changes is $C_{bbb}$, but this is multiplied
with $Y^b$, which vanishes precisely when $C_{bbb}$ jumps.
This type of argument can be used to investigate
the behaviour of various quantities at the transition point
\cite{GreSchShi}. The result for the case at hand is as follows:
\begin{enumerate}
\item
the harmonic functions occuring in the exterior and interior 
solution are the same, i.e., the parameters $c_I$ and $q_I$
are the same.
\item
the scalar fields $Y^I$ are continuously differentiable at $r_{\star}$, 
but the second derivatives are discontinuous.
\item
the function ${\cal V}(Y)$ and, hence, the metric is two-times
continuously differentiable at $r_{\star}$. Higher derivatives
are discontinuous. 
\end{enumerate}
Thus one does not need to introduce a source of stress energy or
charge at $r_{\star}$ in order to continue
the solution.

So far our discussion applies equally to flops and to $SU(2)$
gauge symmetry enhancement. But there is a crucial difference,
which we have to discuss now. For flops the branches $Y^b > 0$ and
$Y^b < 0$ of the extended parameter space are inequivalent, because
the corresponding families of Calabi-Yau spaces are not biholomorphically
equivalent. This is different for $SU(2)$ enhancement,
where $Y^b$ is equivalent to $-Y^b$. In physical terms 
both values are related by an $SU(2)$ gauge transformation, the 
Weyl twist, while in geometrical terms they are related by a
so-called elementary transformation of divisors. 
The triple intersection numbers are related by
\beq
\tilde{C}_{IJK} = C_{MNP} R^M_{\;I} R^N_{\;J} R^P_{\;K}
\eeq
where $(R^{M}_{\;I} ) \in GL(n_V+1,{\bf Z})$ 
satisfies 
$R_{\;I}^{M} R_{\;N}^{I} = \delta^{M}_N$.
As a consequence the theory based on the prepotential
$\tilde{\cal V}(Y)$ can be obtained from the theory defined by
${\cal V}(Y)$ by applying the above transformation. In particular
one can obtain the solutions $Y^I=\tilde{f}^I[\tilde{H}]$ 
of (\ref{FlowEq2}) by transforming
solutions $Y^I=f^I[H]$ of (\ref{FlowEq}):
\beq
\tilde{f}^I[\tilde{H}] = R^I_{\;J} f^J[\tilde{H}\cdot R]  \;.
\eeq
As discussed above, the matching conditions imply that 
the same harmonic functions appear in the interior and
exterior solution. Therefore a solution crossing an $SU(2)$ boundary
takes the form:
\bea
Y^I_{\mscr{(ext)}}(r) &=&  
f^I (H(r)) \;, \;\;\; \mbox{for   }r>r_{\star} \;, \\
Y^I_{\mscr{(int)}}(r) &=& 
R^I_{\;J} f^J(H(r) \cdot R) \;,\;\;\; \mbox{for   } r < r_{\star} \;. 
\eea
For $r< r_{\star}$ the scalars take values in the 
reflected cone $\{R^I_{\;J} Y^J >0 \}$. Since this is 
gauge-equivalent to the original cone $\{Y^I >0 \}$, we can
also parametrize the solution exclusively in terms of scalars
which take values in one of the cones. For example, we can 
apply the Weyl twist to the interior solution, $Y^I \rightarrow
\hat{Y}^I = R^I_{\;J} Y^J$ and obtain:
\bea
Y_{\mscr{(ext)}}^I(r) &=& f^I(H(r)) \;,  \mbox{   for   } r>r_{\star} \;, \\
\hat{Y}_{\mscr{(int)}}^I(r) &=& f^I(H(r)\cdot R) \;,  \mbox{   for   }
r<r_{\star} \;.
\eea
Now the scalars take values inside the original cone for both
the exterior and interior part, but one has to apply a gauge
transformation when the solution reaches $r_{\star}$ in order to connect
the two branches of the solution.

Let us now return to the issue of singularities. Since 
(\ref{FlowEq}) cannot be solved in closed form 
it is difficult to obtain results in a model-independent way. However, 
the study
of a specific family of Calabi-Yau spaces suggests that
solutions cannot become singular as long as the scalars take
values inside the K\"ahler cone. We consider the elliptic fibrations 
over the Hirzebruch surfaces ${\cal F}_0,{\cal F}_1,{\cal F}_2$ 
\cite{LouEtAl}.
Let $Y^I>0$ be the K\"ahler cone and introduce new scalars $S,T,U$
by
\beq
Y^0 = U \;,\;\;\;
Y^1 = T-U \;, \;\;\;
Y^2 = \left\{ \begin{array}{ll}
S-U\;, & \mbox{   for   } {\cal F}_0 \;, \\
S-\ft12(T+U)\;, & \mbox{   for   } {\cal F}_1 \;, \\
S-T \;, & \mbox{   for   } {\cal F}_2 \;. \\ \end{array} \right.
\eeq 
Though the variables $S,T,U$ are not adapted to the K\"ahler cone,
they are convenient, because the prepotentials of all three models
take the same simple form:
\beq
{\cal V} = STU + \ft13 U^3 \;.
\eeq
In this basis the solution of (\ref{FlowEq}) is \cite{KalMohShm}:
\beq
U = \sqrt{ \ft{\Delta}{2}} \;, \;\;\;
T = \sqrt{ \ft{2}{\Delta}} H_S \;, \;\;\;
S = \sqrt{ \ft{2}{\Delta}} H_T \;, \;\;\;
\mbox{where} \;\;\;
\Delta = H_U - \sqrt{ H_U^2 - 4 H_S H_T } \;.
\eeq
One finds that there are two situations where 
(\ref{Rscalar}) becomes singular:
\begin{enumerate}
\item
$H_S(r_{\star}) H_T(r_{\star}) = 0$, which implies 
$U(r_{\star})=0$. $U=0$ is a boundary of the
K\"ahler cone where the volume of some of the two-cycles goes
to zero while the volumes of others diverge.\footnote{
A more generic realization of this
singularity is the collapse of the six-cycle to a four-cycle.
In our model this cannot happen because the total volume
is fixed. This case is not covered by the above list, and no
five-dimensional interpretation is known to the author. However, if one
considers type II string theory on the same Calabi-Yau space, then
$U=0$ corresponds to a non-geometrical Landau-Ginsburg phase.
We thank Albrecht Klemm for drawing our intention to the 
string theory interpretation.} 
The metric on the K\"ahler cone becomes singular and the
five-dimensional effective theory breaks down.  

\item
$H_U(r_{\star})^2 = 4 H_S(r_{\star}) H_T(r_{\star})$.
This does not correspond to a special locus in moduli space, but
the derivatives of the scalars $S,T,U$ diverge whenever the
above equation holds. However, if the scalars take values inside
the K\"ahler cone, then $T>U$, which implies
$H_U > H_S + H_T$. Thus one reaches the boundary $T=U$ 
first and the solution is regular 
at this point.\footnote{
If the solution has been fine-tuned such that 
$T(r_{\star}) = U(r_{\star}) = S(r_{\star})$, then the singularity
occures precisely when the boundary is reached. $T=U=S$ is a 
special point on the boundary where one does not get $SU(2)$
gauge symmetry enhancement, but tensionless strings \cite{KalEtAl}.} 
\end{enumerate}
We conclude that all solutions
are regular as long as the scalars take values inside the K\"ahler cone.
Whether a solution becomes singular when the boundary is reached
depends on the kind of singularity the internal manifold develops.
Note that a solution in the reflected cone is singular if and only if
the gauge-equivalent solution in the original cone is singular.
Since all solutions in the original cone are regular,
it is guaranteed that the continuation of
a black hole solution through $SU(2)$ enhancement yields a 
regular solution. 

We have established for our three models 
that naked singularities are either related
to a breakdown of the effective theory at the boundary 
of the K\"ahler cone or removed by an 
enhan\c{c}on-like mechanism. It is tempting to speculate that this
holds generally. We find it suggestive that the linear 
equation characterising the $SU(2)$ boundary is the
discriminante of a quadratic equation which determines the
presence or absence of a space-time singularity.


It is straightforward to extend the above analysis to 
magnetic black string solutions, which take the form \cite{ChaSab}:
\bea
ds^2 &=& {\cal V}(Y)^{-1/3} ( -dt^2 + dz^2) + {\cal V}(Y)^{2/3} 
dx_{(3)}^2 \;, \nonumber \\
F^I_{mn} &=& - \varepsilon_{mnp} \der_p Y^I \;,\;\;\;
\phi^x = {Y^x}/{Y^0} \;.
\eea
This time the rescaled scalars themselves have to be harmonic,
$\Delta Y^I =0$. Thus the analogue of (\ref{FlowEq}) simply is:
$Y^I = H^I$.
Proceeding in the same way as before one obtains the following 
results:
\begin{enumerate}
\item
naked singularities occure if and only if ${\cal V}(Y(r_{\star}))=0$.
This can only happen on the boundary of the K\"ahler cone. Thus the 
solution is guaranteed to be regular as long as the scalars take 
values inside the K\"ahler cone. Note that
in contrast to black holes we are able to prove this 
for any choice of ${\cal V}(Y)$, because we know the solution explicitly 
in terms of the harmonic functions.  
\item
black string solutions can be continued through flop transitions
and through $SU(2)$ enhancement. The scalars $Y^I$ and the 
gauge fields are smooth at $r_{\star}$, whereas the metric 
is two times continuously differentiable. 
\end{enumerate}

Let us now compare our results to analogous results for the domain 
walls of five-dimensional Horava-Witten theory \cite{OvrEtAl}. Here it was
shown that one can continue the bulk solution through flop transitions 
\cite{GreSchShi}. This is very similar to the above analysis 
for black holes. However, for domain walls the scalars are only
continuous and the metric is only continuously 
differentiable. This reflects that one needs to introduce a 
source of G-flux (though none of stress energy) at the transition 
point. The case of $SU(2)$ enhancement remains to be analysed and
poses new problems, since the $SU(2)$ enhancement is generically
incompatible with the background G-flux. We hope to report on this
question in a future publication. One can also analyse the question
of naked curvature singularities in Horava-Witten domain walls. 
Since the domain wall solution is determined by an algebraic
equation of the form (\ref{FlowEq}), there is a one to one 
correspondence between domain walls and black holes.
It is straightforward to show
that a domain wall exhibits a naked curvature singularity 
if and only if the corresponding black hole solution does. 
This reflects that both space-time geometries are determined
by the same dynamical system on the K\"ahler cone,
as advocated in \cite{BehGukShm}. Also note that the above
analysis of singularities of black holes solutions 
for a particular set of Calabi-Yau
compactifications is identical to the one carried out for the 
corresponding Horawa-Witten domain walls in \cite{KalMohShm}.

Finally we comment on the relation of our results
to the enhan\c{c}on \cite{ClifEtAl}. The point in common 
is that an apparent naked curvature singularity of a 
supergravity solution is absent in M-theory because one first runs
into $SU(2)$ enhancement. The difference is that enhan\c{c}on
geometries generically require to place branes as sources of
charge and stress energy at the transition point. Moreover
the interior solution is typically of a different type than
the exterior one. Our solution seems closest to a particular
type of enhan\c{c}on solutions where the interior solution
is a black hole \cite{Ref?}. It will be interesting to find
the relations of the various incarnations of the enhan\c{c}on mechanism.
We close with speculating that there is a universal mechanism 
in M-theory which avoids space-time singularities by 
the intervention of additional light modes, which are related to
degenerations and topological transitions of the internal space.

{\bf Acknowledgement} I would like to thank the organisers
of the symposium Ahrenshoop 2002 for organizing a stimulating meeting
and for the opportunity to present this work, and I thank
Laur J\"arv,
Albrecht Klemm, Christoph Mayer, Frank Saueressig and Marco 
Zagermann for
useful discussions and comments on the manuscript.

\end{document}